\documentclass[aps,prd,showpacs,floatfix,twocolumn,10pt]{revtex4-1}

\usepackage{epsfig,amssymb,amsfonts,bm}

\def\be{\begin{equation}}
\def\ee{\end{equation}}
\def\bea{\begin{eqnarray}}
\def\eea{\end{eqnarray}}
\def\beas{\begin{eqnarray*}}
\def\eeas{\end{eqnarray*}}

\DeclareMathSymbol{\varGamma}{\mathord}{letters}{"00}

\begin{document}

\title{Remarks on the quantum numbers of $X(3872)$ from the
invariant mass distributions of the $\rho J/\psi$ and $\omega J/\psi$
final states}

\author{C. Hanhart}
\affiliation{Forschungszentrum J\"ulich, Institute for Advanced Simulation, Institut f\"ur Kernphysik (Theorie) and J\"ulich Center for Hadron Physics, D-52425 J\"ulich, Germany}

\author{Yu. S. Kalashnikova}

\author{A. E. Kudryavtsev}

\author{A. V. Nefediev}
\affiliation{Institute for Theoretical and Experimental Physics, 117218,
B.Cheremushkinskaya 25, Moscow, Russia}

\begin{abstract}
We re-analyse the two- and three-pion mass distributions in the decays
$X(3872)\to \rho J/\psi$ and $X(3872)\to \omega J/\psi$ and argue that
the present data favour the $1^{++}$ assignment for the quantum numbers of the
$X$.
\end{abstract}


\maketitle

\section{Introduction}

Since its discovery in 2003 \cite{Xobservation}, the $X(3872)$ charmonium is
subject of many experimental and theoretical efforts aimed at disclosing its
nature---for a recent review see \cite{review}.
The most recent data on the mass of the $X$ is
\cite{Belle2pi}
\be
M_X=(3871.85\pm 0.27({\rm stat})\pm0.19({\rm syst}))~\mbox{MeV},
\label{Xmass}
\ee
with a width of $\varGamma_X<1.2~\mbox{MeV}$. However, the problem of the
quantum numbers for the $X$ is not fully resolved yet: while the analysis of
the $\pi^+\pi^- J/\psi$ decay mode of the $X(3872)$ yields either $1^{++}$ or
$2^{-+}$ quantum numbers \cite{rho,Belle2pi}, the recent
analysis of the $\pi^+\pi^-\pi^0 J/\psi$ mode seems to favour the $2^{-+}$
assignment \cite{Babar3pi}, though the $1^{++}$ option is not excluded.
This question is clearly very central, for the most promising explanations
for the $X$ in the $S$-wave $D\bar D^*$ molecule model \cite{molecule} as well as in the coupled-channel model
\cite{coupled} require the quantum numbers $1^{++}$. In addition, the $X$ cannot be a
naive $c \bar c$ $2^{-+}$ state, for its large branching fraction for the $D^0
\bar D^0 \pi^0$ mode~\cite{DDpi} is not compatible with the
quark-model estimates for the $2^{-+}$ charmonium~\cite{1D2}. So, for the
$2^{-+}$ quantum numbers, very exotic explanations for the $X$ would have to
be invoked.

The aim of the present paper is to perform a combined analysis of the data on
the $\pi^+ \pi^-$ and $\pi^+ \pi^- \pi^0$ mass distribution in the $\pi^+\pi^- J/\psi$ and
$\pi^+\pi^-\pi^0 J/\psi$ mode, respectively. We
find that the $S$-wave amplitudes from the decay of a $1^{++}$ state provide
a better overall description of the data than the $P$-wave ones from the
$2^{-+}$, especially when the parameter range is restricted to realistic
values. We conclude then that the existing data favour
$1^{++}$ quantum numbers of the $X$, however, improved data in the
$\pi^+\pi^-\pi^0 J/\psi$ mode are necessary to allow for definite conclusions
regarding the $X$ quantum~numbers.

\section{Experimental situation}

Recently Belle announced \cite{Belle2pi} the updated
results of the measurements for the reaction $X(3872)\to \pi^+\pi^- J/\psi$:
\beas
{\cal B}_{2\pi}^+&=&[8.61\pm 0.82({\rm stat})\pm 0.52({\rm syst})]\times 10^{-6},\\
{\cal B}_{2\pi}^0&=&[4.3\pm 1.2({\rm stat})\pm 0.4({\rm syst})]\times 10^{-6},
\eeas
for the charged $B^+\to J/\psi\rho K^+$ and neutral $B^0\to J/\psi\rho K^0$ mode, respectively, with ${\cal B}_{2\pi}$ being the product branching fraction ${\rm Br}(B\to KX(3872))\times {\rm Br}(X(3872)\to\pi^+\pi^-J/\psi)$ in the corresponding mode. The number of events in the background-subtracted combined distribution is
$$
N_{2\pi}=196.0^{+18.9}_{-15.2}.
$$
For the decay $X(3872)\to \pi^+\pi^-\pi^0 J/\psi$ Babar reports \cite{Babar3pi}
\beas
{\cal B}_{3\pi}^+&=&[0.6\pm 0.2({\rm stat})\pm 0.1({\rm syst})]\times 10^{-5},\\
{\cal B}_{3\pi}^0&=&[0.6\pm 0.3({\rm stat})\pm 0.1({\rm syst})]\times 10^{-5},
\eeas
for the charged mode $B^+\to J/\psi\omega K^+$ and for the neutral mode $B^0\to J/\psi\omega K^0$, respectively. Similarly to the two-pion case above, ${\cal B}_{3\pi}$ stands for the product branching fraction ${\rm Br}(B\to KX(3872))\times {\rm Br}(X(3872)\to\pi^+\pi^-\pi^0 J/\psi)$ in the corresponding mode. The number of events in the combined distribution is
\be
N^{\rm sig+bg}_{3\pi}=34.0\pm 6.6,\quad N_{3\pi}^{\rm bg}=8.9\pm 1.0,
\ee
and we assume a flat background. Note, the spectrum reported in
\cite{Babar3pi} and used below appears not to be efficiency
corrected. However, since only the shape of this spectrum plays a role for the
analysis (see number-of-event distributions (\ref{N23pi}) below) and we can
reproduce the theoretical spectra of \cite{Babar3pi}, which have
the efficiency of the detector convoluted in via a Monte Carlo simulation,
the invariant mass dependence of the efficiency corrections is expected
to be mild and therefore should not affect our analysis significantly.

Thus the updated ratio of branchings reads \cite{Belle2pi}
\be
\frac{{\cal B}_{3\pi}}{{\cal B}_{2\pi}}=\frac{{\rm Br}(X(3872)\to\pi^+\pi^-\pi^0 J/\psi)}{{\rm Br}(X(3872)\to\pi^+\pi^- J/\psi)}=0.8\pm 0.3.
\label{brratio}
\ee
In our analysis we use the ratio (\ref{brratio}) as well as
\be
N_{2\pi}=196,\quad N_{3\pi}=25.1,
\label{BandN}
\ee
and he corresponding bin sizes are $\Delta E_{2\pi}=20$~MeV and $\Delta E_{3\pi}=7.4$~MeV.

\section{Theoretical $\pi^+\pi^-$ and $\pi^+\pi^-\pi^0$ invariant mass distributions}

As in previous analyses, we assume that the two-pion final state is mediated
by the $\rho$ in the intermediate state, while the three-pion final state is
mediated by the $\omega$. It was shown in \cite{rho} that the
description of the $\pi^+\pi^- J/\psi$ spectrum with the $2^{-+}$ assumption
is improved drastically if the isospin-violating $\rho$-$\omega$ mixing is taken into
account. Theoretical issues of the $\rho$-$\omega$ mixing are
discussed, for example, in \cite{rhoomega}.
Here we include this effect with the help of the prescription used in
\cite{yndurain}, where the transition amplitude is described by the real
parameter $A_{\omega\to\rho}=A_{\rho\to\omega}^\dagger=\epsilon$. Thus, the
amplitudes for the decays $X\to \pi^+\pi^- J/\psi$ and $X\to \pi^+\pi^-\pi^0 J/\psi$  take
the form
\beas
&&A_{2\pi}=A_{X\to J/\psi\rho}G_\rho A_{\rho\to 2\pi}\\
&&\hspace*{3cm}+A_{X\to J/\psi\omega}G_\omega \epsilon G_\rho A_{\rho\to2\pi},\\
&&A_{3\pi}=A_{X\to J/\psi\omega}G_\omega A_{\omega\to 3\pi}\\
&&\hspace*{3cm}+A_{X\to J/\psi\rho}G_\rho \epsilon G_\omega A_{\omega\to3\pi},
\eeas
where the vector meson propagators are
\be
G_V^{-1}=m_V^2-m^2-im_V\varGamma_V(m),\quad V=\rho,\omega,
\ee
with $m=m_{\pi\pi}$ ($m=m_{\pi\pi\pi}$) being the $\pi^+\pi^-$
($\pi^+\pi^-\pi^0$) invariant mass in the $\pi^+\pi^- J/\psi$ ($\pi^+\pi^-\pi^0 J/\psi$) final state. Masses of the vector mesons used below are \cite{PDG}
$$
m_\rho=775.5~\mbox{MeV},\quad m_\omega=782.65~\mbox{MeV}.
$$
The complex mixing amplitude multiplying the $\omega$ propagator used, for example, in
\cite{rho} to analyse the two-pion spectrum,
in our notation reads as $\epsilon G_\rho(m_\omega)$; in particular we reproduce naturally
the phase of 95$^{\rm o}$ quoted in \cite{rho}.
Note, as we shall only study the invariant
mass distributions of the two final states, we do not need to keep explicitly
the vector nature of the intermediate states.
For the ``running'' $\rho$ meson width we use
$$
\varGamma_\rho(m)\approx\varGamma_{\rho\to2\pi}(m)=\varGamma_\rho^{(0)}\frac{m_\rho\;q(m)}{m\;q(m_\rho)}\left[\frac{f_{1\rho}(q(m))}{f_{1\rho}(q(m_\rho))}\right]^2,
$$
where $q(m)=\sqrt{m^2-4 m_\pi^2}/2$, $f_{1\rho}(q)=(1+r_\rho^2 q^2)^{-1/2}$, with $r_\rho=1.5$~GeV$^{-1}$ and with the nominal $\rho$ meson width $\varGamma_\rho^{(0)}=146.2$~MeV.

For the $\omega$ meson ``running'' width (the nominal width being $\varGamma_{\omega}^{(0)}=8.49$~MeV), the $3\pi$ and $\pi\gamma$ decay modes are summed, with the branchings
\be
{\rm Br}(\omega\to 3\pi)=89.1\%,\quad {\rm Br}(\omega\to\pi\gamma)=8.28\%.
\ee

In particular,
\be
\varGamma_{\omega\to\pi\gamma}(m)=\varGamma_{\omega\to\pi\gamma}^{(0)}\left[\frac{m_\omega (m^2-m_\pi^2)}{m (m_\omega^2-m_\pi^2)}\right]^3,
\ee
while, for the $\varGamma_{\omega\to3\pi}(m)$, we resort to the expressions derived in \cite{kuraev}, with a reduced contact term which provides the correct nominal value of the $\omega\to3\pi$ decay width~\cite{rajeev}.

The transition amplitudes for the decays $X\to J/\psi V$ are parameterised
in the standard way, namely,
\be
A_{X\to J/\psi V}=g_{XV}f_{lX}(p),
\ee
with the Blatt-Weisskopf ``barrier factor''
\be
f_{0X}(p)=1,~\mbox{and}~f_{1X}(p)=(1+r^2 p^2)^{-1/2},
\ee
for the $1^{++}$ and $2^{-+}$ assignment, respectively.  Here $p$ denotes the
$J/\psi$ momentum in the $X$ rest frame.  The ``radius'' $r$ is not well
understood. If one associates it with the size of the $X\to \rho(\omega)J/\psi$ vertex, it might be related to the range of force. In the
quark model this radius is $r\sim 0.2~\mbox{fm}=1~\mbox{GeV}^{-1}$.  This is
also in line with the inverse mass of the lightest exchange particle allowed
between $J/\psi$ and $\rho/\omega$, namely, $f_0(980)$. On the other hand, a
larger value $r=5$~GeV$^{-1}$ is used in the experimental analysis of
\cite{Belle2pi}. Therefore, in the analysis presented below we use both values
$r=1$ GeV$^{-1}$ as well as $r=5$ GeV$^{-1}$, keeping in mind that smaller
values of $r$ are preferred by phenomenology.

The theoretical invariant mass distributions for the $\pi^+\pi^-$ and $\pi^+\pi^-\pi^0$ final state take the form:
\bea
\frac{d{\rm Br}_{2\pi}}{dm}={\cal B}m_\rho\varGamma_{\rho\to 2\pi} p^{2l+1}f_{lX}^2(p)\left|R_XG_\rho+\epsilon G_\rho G_\omega\right|^2,\nonumber\\[-2mm]
\label{Br23pi}\\[-2mm]
\frac{d{\rm Br}_{3\pi}}{dm}={\cal B}m_\omega\varGamma_{\omega\to3\pi} p^{2l+1}f_{lX}^2(p)\left|G_\omega+\epsilon R_X G_\omega G_\rho \right|^2,\nonumber
\eea
where $R_X=g_{X\rho}/g_{X\omega}$ and the parameter ${\cal B}$ absorbs the details of the short-ranged dynamics
of the $X$ production.

The theoretical number-of-event distributions read
\bea
N_{2\pi}(m)=\frac{N_{2\pi}\Delta E_{2\pi}}{B_{2\pi}}\times\frac{d{\rm Br}_{2\pi}}{dm},\nonumber\\[-2mm]
\label{N23pi}\\[-2mm]
N_{3\pi}(m)=\frac{N_{3\pi}\Delta E_{3\pi}}{B_{3\pi}}\times\frac{d{\rm Br}_{3\pi}}{dm}.\nonumber
\eea

The $\rho$-$\omega$ mixing parameter $\epsilon$ is extracted from the $\omega\to2\pi$ decay width (${\rm Br}(\omega\to 2\pi)=1.53\%$). The corresponding amplitude reads
\be
A_{\omega\to2\pi}=\varepsilon G_\rho A_{\rho\to2\pi},
\ee
and we find that
\be
\epsilon\approx\sqrt{m_\omega m_\rho\varGamma_\rho\varGamma_{\omega\to2\pi}}\approx 3.4\times 10^{-3}~\mbox{GeV}^2.
\ee

\begin{table*}[t]
\caption{Sets of parameters and the quality of the combined fits to the
data. The last column refers to the presentation of the different fits in
Fig.~\ref{fig1}.}
\label{t1}
\begin{ruledtabular}
\begin{tabular}{c|ccccccc}
Fit&Wave&$J^{PC}$&$r$, GeV$^{-1}$&$R_X=g_{X\rho}/g_{X\omega}$&$\chi^2/N_{\rm dof}$&CL&
Curve/Markers in Fig.~\ref{fig1}\\
\hline
$S$&$S$&$1^{++}$&$-$&$0.26^{+0.08}_{-0.05}$&1.07&37\%&Solid (red)/Triangles\\
$P_{5}$&$P$&$2^{-+}$&5&$0.15^{+0.05}_{-0.03}$&1.33&14\%&Dash-dotted (blue)/Squares\\
$P_{1}$&$P$&$2^{-+}$&1&$0.09^{+0.03}_{-0.02}$&2.77&$10^{-5}$&Dashed (green)/Diamonds
\end{tabular}
\end{ruledtabular}
\end{table*}
\vspace*{-6mm}

\section{Fitting strategy and results}

The number-of-event distributions (\ref{N23pi}) possess 3 free parameters: the
``barrier'' factor $r$, the ratio of couplings $R_X$ and the overall normalisation parameter ${\cal B}$.
As outlined above, we perform the
analysis for two values of $r$, namely, the preferred value of 1 GeV$^{-1}$
and a significantly larger value of 5 GeV$^{-1}$ used in earlier analyses.
Since the
normalisation factor ${\cal B}$ drops out from the ratio of the two
branchings, we extract the ratio
$R_X$ directly from the integrated data, that is from the relation
\be
\left(\int \frac{d{\rm Br}_{3\pi}}{dm}dm\right)\left/\left(\int \frac{d{\rm Br}_{2\pi}}{dm}dm\right)\right.={\cal B}_{3\pi}/{\cal B}_{2\pi},
\ee
where the value of the ratio on the right-hand side is fixed by Eq.~(\ref{brratio}), and in the integration
above we have cut off the $\pi^+\pi^-$ invariant mass at 400 MeV, as in \cite{Belle2pi}, and the
$\pi^+\pi^-\pi^0$ invariant mass at 740 MeV, as in \cite{Babar3pi}.
Therefore the norm ${\cal B}$ is our only fitting parameter which governs the
overall strength of the signal in both channels simultaneously, while the
shape of the curves is fully determined from other sources.

\begin{center}
\begin{figure*}[ht]
\centerline{\epsfig{file=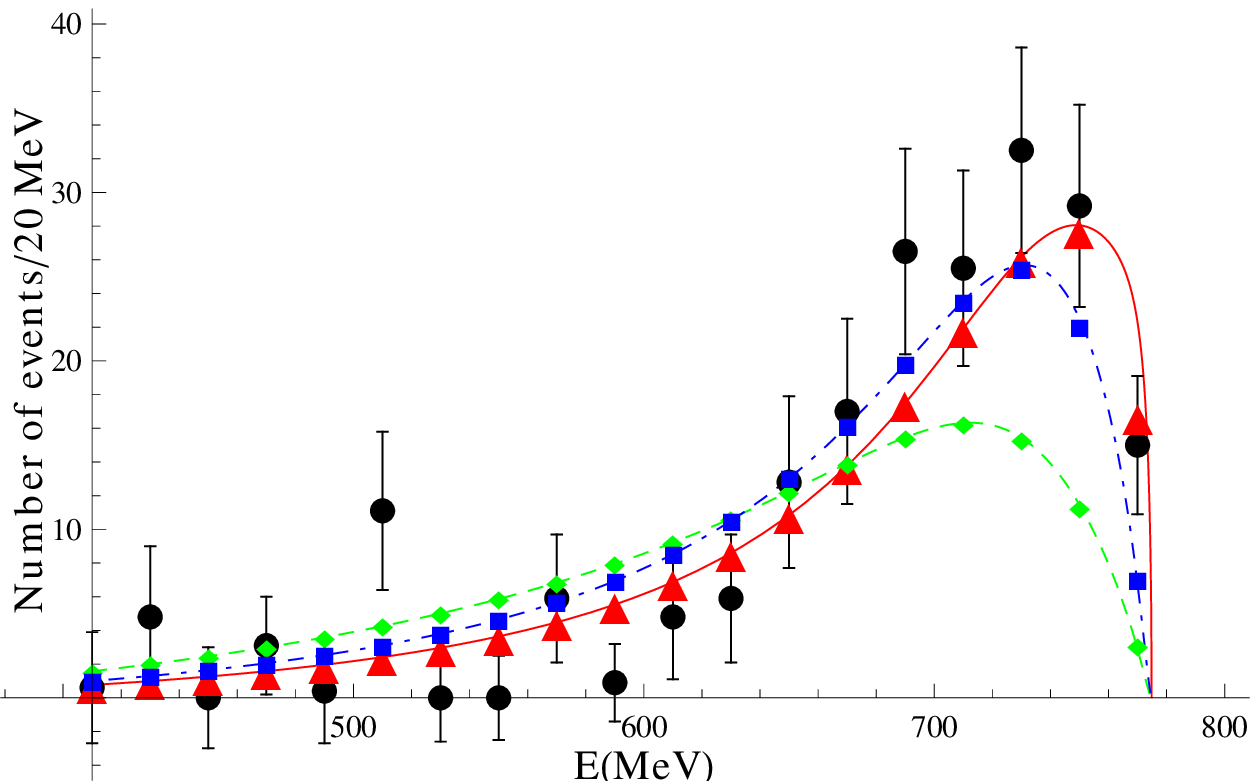, width=9cm}\hspace*{1mm}\epsfig{file=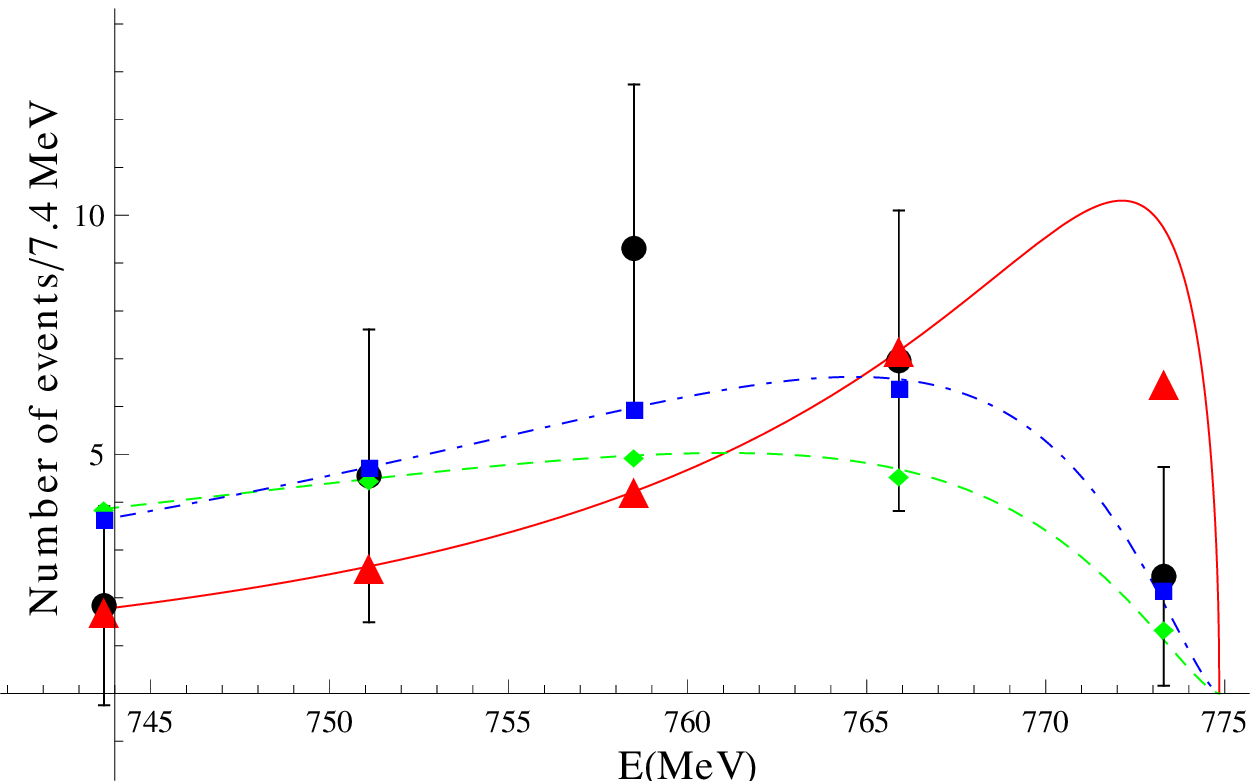, width=9cm}}
\caption{Comparison of the theoretical predictions with the data. Experimental
data, taken from \cite{Belle2pi,Babar3pi} for the two-pion (left plot) and three-pion (right plot) spectrum, respectively, are given by black circles with error bars, while the theoretical
results are shown both as the continuous number-of-event distributions
(\ref{N23pi}) as well as by markers for the corresponding integrals over the bins. Fit $S$ is
shown as the (red) solid line and triangles, fit $P_1$ is shown as the
(green) dashed line and diamonds, while fit $P_5$ is shown as the (blue)
dash-dotted line and squares. See Table~\ref{t1} for the
details.}\label{fig1}
\end{figure*}
\end{center}

In Table~\ref{t1}, we list the parameters of the 3 combined fits to the
data, found for the 2 values of the Blatt-Weisskopf parameter $r$. The
corresponding line shapes and the result of the integration in bins are
shown in Fig.~\ref{fig1}.

One can see from Table~\ref{t1} and Fig.~\ref{fig1} that the best overall
description of the data for the two channels under consideration is provided
by the $S$-wave fit. The $P$-wave fit is capable to provide the description of
the data of a comparable (however somewhat lower) quality, only for large
values of the Blatt-Weisskopf parameter $r$, $r=5$~GeV$^{-1}$. The $P$-wave
fit becomes poorer when the Blatt-Weisskopf parameter is decreased, and for
values of $r$ of order 1~GeV$^{-1}$, the quality of the $P$-wave fit is
unsatisfactory, which is the result of a very poor description of the two-pion
spectrum---see the dashed (green) curve in Fig.~\ref{fig1}. Varying the ratio of
branchings ${\cal B}_{3\pi}/{\cal B}_{2\pi}$ around its central value within
the experimental uncertainty interval [see Eq.~(\ref{brratio})] leads only to
minor changes in the fits and  does not affect the
conclusions.
\vspace*{-10mm}

\section{Discussion}

Since no charged partners of the $X(3872)$ are observed
experimentally, it is supposed to be (predominantly) an isoscalar. Then
the ratio $R_X=g_{X\rho}/g_{X\omega}$ measures the strength of the isospin violation
in the $X \to V J/\psi$ decay vertex.
As discussed above, this ratio is extracted directly from the data
on the ratio of the branchings (\ref{brratio}).

An isospin-violating observable for a compact charmonium is
the ratio of the branching fractions for the $\psi(2S)$ decays into $\eta J/\psi$
and $\pi^0 J/\psi$ final states as
\be
R_{\psi(2S)}=\frac{g_{\pi^0 J/\psi}}{g_{\eta J/\psi}}=
\sqrt{\frac{{\rm Br}(\psi(2S) \to \pi^0 J/\psi)}{{\rm Br}(\psi(2S) \to \eta J/\psi)}
\frac{k_{\eta}^3}{k_{\pi^0}^3}} \approx 0.03,
\label{rcharm}
\ee where $k_{\pi^0}$ and $k_{\eta}$ are the center-of-mass momenta of $\pi^0$
and $\eta$, respectively, and the $\psi(2S)$ branching fractions are taken
from \cite{PDG}. However,  since
here also the denominator violates a symmetry, namely, SU(3), and there might
be significant meson-loop contributions \cite{psiloops}, the estimate
(\ref{rcharm}) is to be regarded as a conservative upper bound for the isospin violation strength
for compact charmonia.

In contrast to this,
in the $S$-wave molecular picture for the $X$,
iso\-spin violation is enhanced significantly compared to the strength (\ref{rcharm}) for it
proceeds via intermediate $D\bar{D}^*$ states and is
therefore driven by the mass difference $\Delta \approx 8$~MeV of the neutral and
charged $D\bar{D}^*$ threshold---see, for example,
\cite{suzuki,oset}. An order-of-magnitude estimate is provided by the
expression
\be
R_X^{\rm mol}\sim\left|\frac{I_0(M_X)-I_c(M_X)}{I_0(M_X)+I_c(M_X)}\right| \sim
\frac{\sqrt{m_D\Delta}}{\beta} \sim 0.13,
\label{Rmol}
\ee
where $m_D$ is the $D$ meson mass, while $I_0(M_X)$ and $I_c(M_X)$
denote the amplitudes corresponding to loop diagrams with neutral and charged $D\bar D^*$
intermediate states, respectively, evaluated at the $X$
mass. They are composed of two terms, the strongly channel-dependent analytic continuation of the unitarity cut, proportional to the
typical momentum of the meson pair, and the weakly channel-dependent principle
value term, whose size is identified with the inverse range of forces of order
of 1~GeV (see above). This estimate is within a factor of 2 consistent with the value
$R_X\sim 0.26$ found from our $S$-wave fit---see Table~\ref{t1}.

On the other hand, if the $X$ has the quantum numbers $2^{-+}$, one should
expect the isospin violation in the $X$ wave function to be of the natural
charmonium size, and thus of the order of $10^{-2}$---see discussion below
Eq.~(\ref{rcharm}), since the $D\bar D^*$ loop effects are suppressed by the
additional centrifugal barrier: the estimate analogous to
Eq.~(\ref{Rmol}) now reads $R_X^{\rm mol}\sim(\sqrt{M_D\Delta}/\beta)^3 \sim
2\times 10^{-3}$. Thus, for the state with the quantum numbers $2^{-+}$, one
expects values of at most $R_X\sim 10^{-2}$, that are significantly smaller than those
following from the data (see Table~\ref{t1}).  One is led to conclude
therefore that for $R_X\gtrsim 0.1$, needed for the quantum numbers $2^{-+}$
to be consistent with the data on the $X$ decays, a new, yet unknown,
isospin violation mechanism would have to be invoked.

\section{Conclusions}

We conclude therefore that, although the present quality of the data in the
$X\to \pi^+\pi^-\pi^0 J/\psi$ channel is not sufficient to draw a definite
conclusion concerning the quantum numbers of the $X(3872)$, the combined
analysis of the existing two- and three-pion spectra favours the $S$-wave fit,
related to the $1^{++}$ assignment for the $X(3872)$, over the $P$-wave fit,
related to the $2^{-+}$ assignment. We notice that an acceptable $P$-wave fit calls
for a large range parameter in the Blatt-Weisskopf form factor which meets certain difficulties with its phenomenological interpretation.
In addition, while the value $R_X=g_{X\rho}/g_{X\omega}$ can be understood
theoretically for the $1^{++}$ assignment, the value extracted for the
$2^{-+}$ assignment is too large to be explained from known mechanisms of the
isospin violation.
\begin{acknowledgments}
We acknowledge useful discussions with E.~Braaten, S. Eidelman, F.-K. Guo, and R. Mizuk. The work was supported in parts by funds provided from the Helmholtz
Association (Grant Nos VH-NG-222 and VH-VI-231), by the DFG (Grant Nos SFB/TR 16
and 436 RUS 113/991/0-1), by the EU HadronPhysics2 project, by the RFFI (Grant Nos RFFI-09-02-91342-NNIOa and
RFFI-09-02-00629a), and by the State Corporation of Russian Federation ``Rosatom.''
\end{acknowledgments}

\end{document}